\documentclass[preprint2,]{aastex}
\begin{document}
\title{Experimental Study on Bouncing Barriers in Protoplanetary Disks}
\author{T. Kelling, G. Wurm and M. K{\"o}ster}
\affil{Faculty of Physics, University Duisburg-Essen, Lotharstr. 1, 47057 Duisburg, Germany}
\email{thorben.kelling@uni-due.de}

\begin{abstract}
For dust aggregates in protoplanetary discs a transition between sticking and bouncing in individual collisions at mm to cm size has been observed in the past. This lead to the notion of a bouncing barrier for which growth gets stalled. Here, we present long term laboratory experiments on the outcome of repeated aggregate collisions at the bouncing barrier. About 100 SiO$_2$ dust aggregates of 1 mm in size were observed interacting with each other. Collisions occured within a velocity range from below mm/s up to cm/s. Aggregates continuously interacted with each other over a period of 900 s. During this time more than $10^5$ collisions occured. Nearly 2000 collisions were analyzed in detail. No temporal stable net growth of larger aggregates was observed even though sticking collision occur. Larger ensembles of aggregates sticking together are formed but were disassembled again during the further collisional evolution. The concept of a bouncing barrier supports the formation of planetesimals by seeded collisional growth as well as by gravitational instability favouring a significant total mass being limited to certain size ranges. Within our parameter set the experiments confirm that bouncing barriers are one possible and likely evolutionary limit of a self consistent particle growth.
\end{abstract}

\maketitle

\section{Introduction}
\label{sec:intro}

There has been tremendous progress in recent years concerning the assembly
of km-size planetesimals as early phase of planet formation. In a first step dust grains
couple well to the gas in protoplanetary discs and collide gentle enough that sticking and growth of (initially fractal) dust aggregates is warranted \citep{chokshi1993,dominik1997,wurm1998,blum2000,blum2000b,paszun2006,blum2008,wada2009,birnstiel2012}. In some cases charging might prevent collisions as suggested by \citet{okuzumi2009}. In others, charges accelerate the process \citep{konopka2005}. However, in the dense midplane charging is unlikely to occur and growth of mm-size aggregates is undisputed.

There is little doubt that very porous, fractal aggregates evolve initially. However, the energy eventually is enough to restructure particles. With micrometer silicate grains this occurs at cm-size as experimentally shown by Blum and Wurm (2000) at low speed. While the aggregates which get compacted now are still initially highly porous they are compacted by further collisions even at low speed far below 1 m/s. This has been shown by Weidling et al. (2009) which find in laboratory experiments that a filling factor of 0.36 results after a large number of collisions of an individual mm-size dust 
aggregate. As detailed below this matches the volume filling factor used in the experiments reported here. For this later stage \citet{zsom2010} proposed a bouncing barrier for this size range based on experimental results \citep{blum2008,heisselmann2010}. The basic physics behind the bouncing is that dust aggregates have been restructured by previous collisions and are more or less compact now. As collision velocities are still well below a m/s no significant energy dissipation by further restructuring is possible. The collisions are essentially elastic and rebound is likely. Bouncing has been studied in experiments \citep{weidling2012} and numerical simulations e.g. by \citet{wada2011} or \citet{seizinger2013} and bouncing is also present in experiments with larger solid impactors at low velocities \citep{colwell2003,colwell2008}. This paper is focused on this stage of pre-planetesimal evolution and we do not consider the initial collisions of very fluffy mm-dust aggregates of low filling factor as we explicitly perform experiments to probe the proposed bouncing barrier.

The behaviour of water ice as solid beyond the snowline might be somewhat different but sticking, bouncing (and fragmentation) are also observed and simulated for ice particles \citep{supulver1995, higa1998, schaefer2007, wada2009, okuzumi2012}. 
This might shift the bouncing barrier to larger sizes. Sublimation, condensation and sintering are
important processes here which might change the setting quite a bit though \citep{saito2011,sirono2011,aumatell2011,ros2013}.
We only consider refractory dust in our experiments reported here.

Seeming to be an obstacle to planetesimal formation, the bouncing barrier for dust aggregates of a certain size (mm to cm) 
might be an important milestone for planetesimal formation.
There are two possibilities how to proceed from here on the way to planetesimals. 

If the gas densities or the location in the disc are right turbulence, stable eddies or streaming instabilities might concentrate solids which then
get bound gravitationally and support the formation of larger objects rather fast \citep{johansen2007,cuzzi2008,chiang2010,dittrich2013}. These studies, so far, do not include collisional physics. If conditions are less favourable larger initial particle sizes (e.g. decimeter) are needed
in large amount but mm-sizes might be sufficient in some cases. The rapid formation
of planetesimals within a few orbital timescales would also decrease the problem (or even use it in a streaming instability) that solid particles drift inwards. This
radial drift estimated to be as large as 1 AU in 100 years has been a Damokles' 
sword for a long time \citep{weidenschilling1977}. The problem might also be solved
by considering more complex or realistic disc models with pressure bumps 
or disc edges in different locations where particles would be concentrated
\citep{barge1995,klahr1997,brauer2008,pinilla2012,ayliffe2012}.

The second way to proceed from mm-particles to planetesimals is 
still a model of collisional growth. This is
also not without further assumptions to be detailed in future work, but 
\citet{windmark2012} showed that 
some seeds of larger size are enough to promote growth. This is based on the fact
that collisions with larger objects get faster (tens of m/s) and accretion is now
possible again as seen in a series of experiments \citep{wurm2005,teiser2009b,kothe2010,teiser2011}. 
This requires some particles to jump over the bouncing
barrier and act as seeds for further growth. \citet{windmark2012b} proposed that 
slow collisions in a turbulent disc - unlikely as they might be - could provide a 
small number of seeds. This might go together with results from microgravity
experiments that once a number of compact aggregates form a new aggregate
this allows further growth over some velocity range \citep{weidling2012,kothe2013}. From a set of different laboratory experiments \citet{jankowski2012}  found that sticking probablities are enhanced if the aggregates are composed of grains of 10$\mu m$ in size
instead of smaller grains. These are on the
upper end of reasonable grain sizes in discs but might - even if
in small number - seed further growth. 

Whatever the process to form planetesimals, eventually, important is the fact that the bouncing barrier 
is actually needed or at least very beneficial for the further evolution. If all particles could grow the process
would stall in a collisional model as especially at larger size,
collisions between equal size aggregates will destroy particles even at low collision speeds. This has e.g. been seen in experiments on dust
aggregate collisions from cm to dm by \citet{beitz2011}, \citet{schraepler2012} and \citet{deckers2013}.
With small particles below the bouncing barrier being present, a small number of large objects can
grow on their account. If no bouncing (or fragmentation) barrier would exist too little
particles might be present in the critical size range of Stokes number 1 needed
in the gravitoturbulent or streaming instability models. 

Either way, bouncing of a significant part of i.e. mm-aggregates -- though counterintuitive as it might be -- is likely a key process to form larger bodies.

So far, the bouncing barrier had only been studied in individual collisions.
Small probabilities for sticking two millimeter aggregates together exist and
give the idea that continuous growth might be possible \citep{weidling2012,jankowski2012,kothe2013}. However, detachment in a further collision might also prevent further growth as the connection between two compact dust aggregates is very weak \citep{jankowski2012}. In a somewhat different setting in Saturn's rings but putting emphasize on weak connections particle release is also discussed \citep{bodrova2012}. Therefore, the study of individual collisions is not enough to answer if further growth is possible. In this paper we study the long term collisional evolution of a large number of mm-size aggregates in a  laboratory experiment. We show that the bouncing barrier is indeed a robust part of collisional evolution in protoplanetary discs.

\section{Experiments}
\label{sec:setup}

\begin{figure}
\centering
\includegraphics[width=84mm]{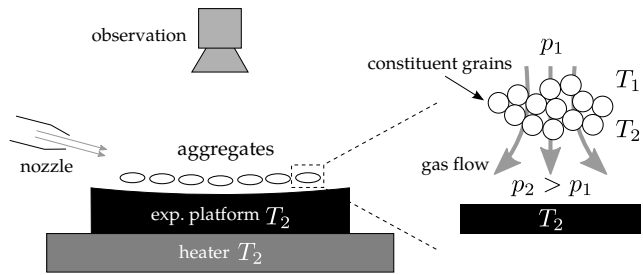}
\caption{The experimental setup is placed within a vacuum chamber. On a slightly concave platform with a heater below are $\sim $100 flat, cylindrically shaped SiO$_2$ aggregates. A camera observes the ensemble from above and a gas flow through a nozzle can be used to enhance the relative velocity between the aggregates. The aggregates are levitated as they become Knudsen compressors (see text for details).}
\label{fig:setup}
\end{figure}

\begin{figure}
\centering
\includegraphics[width=84mm]{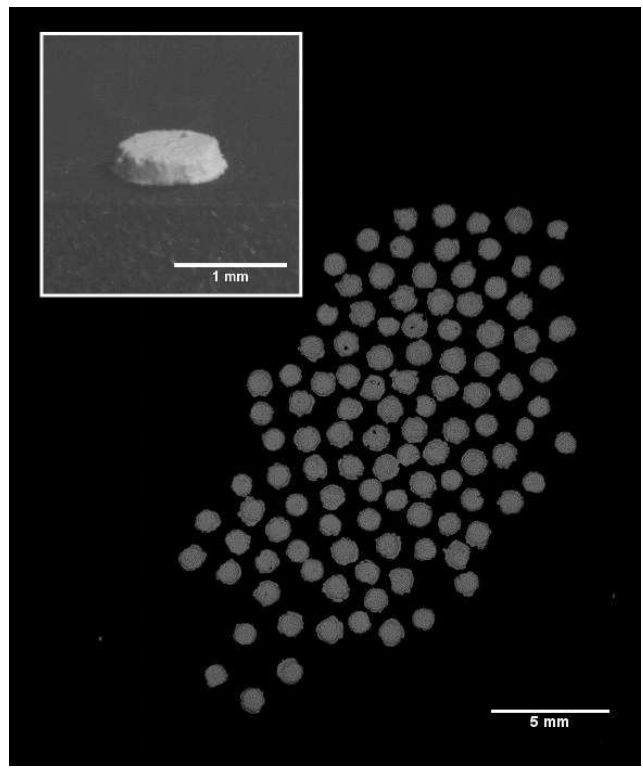}
\caption{SiO$_2$ aggregate sample as observed by the high speed camera. The inset is a microscope image of one of the aggregates.}
\label{fig:agg}
\end{figure}

It was only recently discovered that sub-mm to cm size dust and ice aggregates  as highly porous objects can be levitated in a temperature gradient field at low ambient pressure over a smooth surface \citep{kelling2009,kelling2011c,aumatell2011,jankowski2012}. The lift is generated as the particles act like Knudsen compressors (see below).  \cite{kelling2009} and \cite{jankowski2012} showed, that large numbers of aggreates are easily levitated at the same time. The composition can vary, e.g aggregates from SiO$_2$, basalt or graphite powder with a broad size distribution might be levitated. We used this experimental technique here to generate over 100 free moving mm sized  SiO$_2$ (quartz) aggregates. This material was used frequently in past
experiments. The consitituent grains are between $0.1-10$ $\mu$m in size with  80\% of the grains between $1-5$ $\mu$m (manufacturer: Sigma Aldrich, Tab. \ref{tab:coll}). The basic experimental setup is depicted in Fig.\ref{fig:setup}.
The experimental setup is placed within a vacuum chamber. A heater is coupled to a slightly concave and black platform (radius 30 mm, center depth 0.8 mm) on which $\sim$100 flat cylindrical SiO$_2$ dust aggregates are placed. The aggregates used in the experiments are shown in Fig.\ref{fig:agg}.
The slightly concave shape of the experimental platform  ensures that the levitating aggregates are gently forced to move towards the center of the platform which prevents them from leaving the platform but forces them to interact with each other. A nozzle is installed in some cm distance from the levitating aggregates to excite the aggregate essemble through an air flow. With this gas flow the collision velocity of the aggregates can be enhanced. A camera with 200 frames per second is used to study the details of the motion and the collisions of the aggregates (high speed recording phase) and for the long term recording a second camera takes an image of the set every 3 seconds over a total time period of 15 minutes (long time recording phase). The ambient gas pressure is set to $p= 15 \pm 1$ mbar and the heater temperature is set to $T_2\simeq 800$ K. 

The mechanism behind the levitation of the aggregates is a temperature driven gas flow (thermal creep) through the dust aggregates and a resulting pressure build up below (Knudsen compressor). \cite{knudsen1909} showed that a gas flow is generated by temperature gradients which can establish an overpressure. In his experiments \cite{knudsen1909} connected two gas reservoirs at different temperatures $T_1$ and $T_2>T_1$ by a tube with diameter $d_t$ small compared to the mean free path $\lambda$ of the gas molecules ($d_t\ll \lambda$). In equilibrium the pressure $p_1$ and $p_2$ in the two chambers are related by
\begin{equation}
\frac{p_2}{p_1} =\sqrt{\frac{T_2}{T_1}}.
\end{equation}
With a cascade of several connected chambers \cite{knudsen1909} reached a compression ratio of $p_2/p_1\simeq 10$ between the first and the last chamber. \cite{muntz2002} showed, that the overpressure $\Delta p = |p_1-p_2|$ at intermediate Knudsen numbers $Kn$ in the warmer chamber is
\begin{equation}
\Delta p = p_{avg}\frac{\Delta T}{T_{avg}}\frac{Q_T}{Q_P}.\label{eq:muntz}
\end{equation}
It is $p_{avg} = (p_1+p_2)/2$ the average pressure, $\Delta T = |T_1-T_2|$ the temperature difference between the two gas reservoirs, $T_{avg} = (T_1+T_2)/2$ the average temperature and $Q_T/Q_P$ the the ratio of the coefficients of the temperature induced gas flow ($Q_T$) and the back flow of the gas ($Q_P$). Values for $Q_T/Q_P$ are given by \cite{muntz2002}. The Knudsen number $Kn = \lambda/d$ is definded as the ratio of the mean free path of the gas molecules to a relevant geometric length $d$.

Dust aggregates are not solid bodies but have pores at approximately the same size as their constituent grains \citep{jankowski2012}. These pores form channels which connect the gas volume above the aggregate with the gas volume below the aggregate. Hence, an aggregate can be interpreted as a collection of microchannels. The heater heats the bottom of the aggregates. The heat is transported by thermal conductivity through the aggregate towards the top where thermal radiation cools the aggregate. A temperature difference between the bottom and the top is established and gas flows through the aggregate (thermal creep). Depending on the thermal conductivity $\kappa_{agg}$ and the thickness (height $h$) of the aggregate a more or less prominent temperature difference $\Delta T$ is established between the inlet of the micro-channels (top side, $T_1<T_2$) and the outlet (bottom side, $T_2>T_1$). The temperatures $T_1$ and $T_2$ are related by (neglecting the ambient temperature)
\begin{equation}
\sigma T_1^4 = \frac{\kappa_{agg}}{h}(T_2-T_1).
\end{equation}
According to Eq.(\ref{eq:muntz}) an overpressure is created below the aggregate. If the overpressure is sufficient, the aggregates are levitated. While levitating, the gas which is compressed  by thermal creep through the aggregates escapes through the open sides below the aggregates (see also Fig.\ref{fig:setup}). As we use dust with a similar grain size distribution, similar ambient gas pressure and similar temperatures as \cite{kelling2009} and \cite{jankowski2012} we have the following values: the thermal conductivity of the aggregates ($\kappa_{agg}\simeq 0.1$ W/(m K)), the temperature difference over the aggregates ($\Delta T \simeq 20$ K) and the resulting overpressure below the aggregates $\Delta p\simeq 5$ Pa. The ratio of the force lifting the aggregate induced by the overpressure $F_{\Delta p}$ to the gravitational force $F_G$ is then $F_{\Delta p}/F_G \simeq 5$. 

\begin{figure}
\centering
\includegraphics[width=84mm]{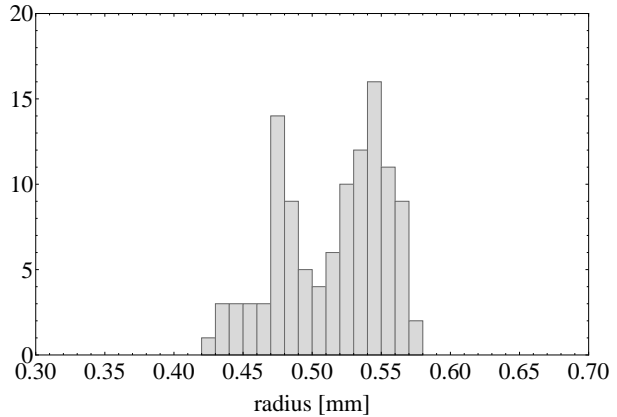}
\caption{Size distribution of the aggregates. The radii $r$ are 2d equivalent radii calculated from the imaged cross section $A$ of an aggregate  ($r=\sqrt{A/\pi}$).}
\label{fig:size}
\end{figure}

\begin{figure}
\centering
\includegraphics[width=84mm]{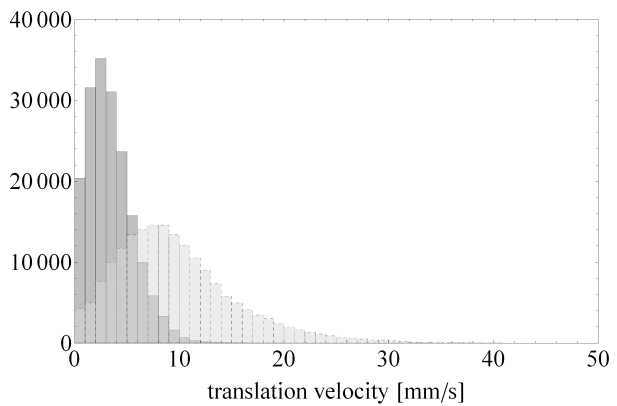}
\caption{Translation velocity distribution of the aggregates with (light grey, dashed) and without excitation (dark grey, solid). The translation velocity is determined from frame to frame in the high speed recording phase.}
\label{fig:vel1}
\end{figure}

To allow a contact free levitation the aggregates have to have a smooth bottom side. Also to avoid any ambiguity due to different size, mass or shape we did not choose particle aggregates randomly but prepared them in a specific procedure. The dust was filled into
moulds within a thin steel plate, manually compressed and any excess dust was scraped off along the top of the plate.  The dust was then knocked out of their moulds resulting in the given flat cylindrical aggregates with smooth top and bottom sides.
The mean of the size distribution of the aggregates radii is at 515 $\mu$m (Fig.\ref{fig:size})
We carried out collision experiments with and without excitation by minor amounts of air through a nozzle. The velocity distributions of the aggregates is shown in Fig.\ref{fig:vel1}.

\section{Collisions}
\label{sec:results}

\begin{figure}
\centering
\includegraphics[width=84mm]{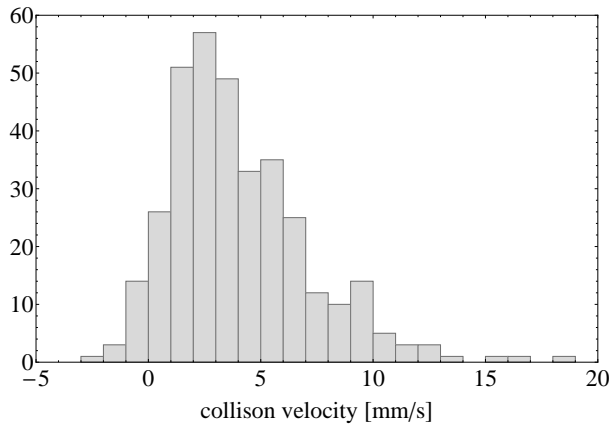}
\caption{Collision velocity distribution without excitation.  A negative velocity means that the two aggregates move away from each other but collide due to rotation.}
\label{fig:vcoll1}
\end{figure}

\begin{figure}
\centering
\includegraphics[width=84mm]{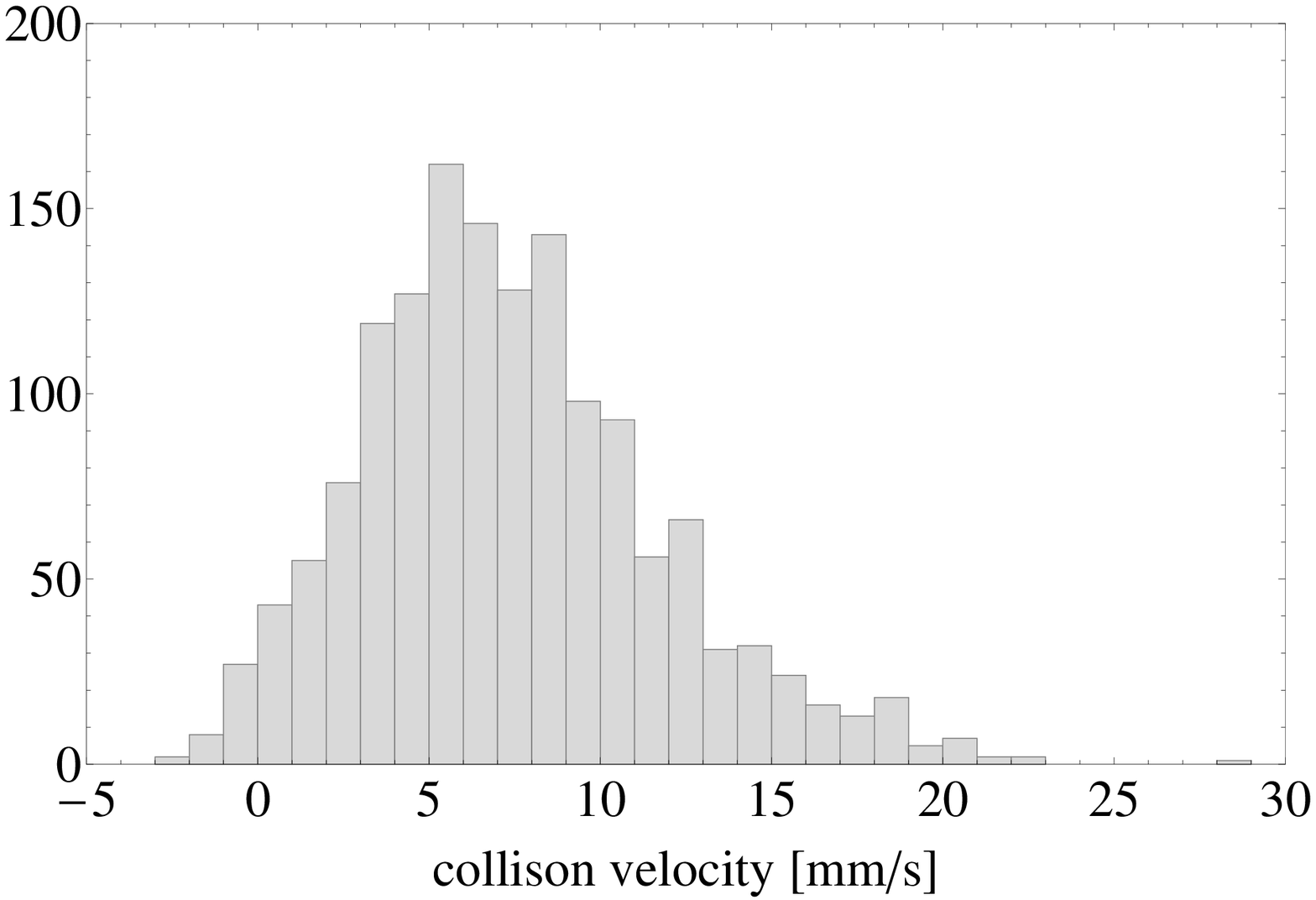}
\caption{Collision velocity distribution with excitation. A negative velocity means that the two aggregates move away from each other but collide due to rotation.}
\label{fig:vcoll2}
\end{figure}

Tab.\ref{tab:coll} lists the details of the aggregates and collision events gained through the high speed recording phase and the extrapolated events for the long term recording.
\begin{table}
\centering
\caption{Aggregate details and collision events.}
\label{tab:coll}
\begin{tabular}{lrr}
\hline
eq. radius $\bar{r}_{agg}$&\multicolumn{2}{r}{$(515\pm 40)$ $\mu$m}\\
height $h_{agg}$&\multicolumn{2}{r}{$(200\pm 30)$ $\mu$m}\\
mass $m_{agg}$&\multicolumn{2}{r}{$(2.13\pm 0.05)\times 10^{-7}$ kg}\\
density $\rho$&\multicolumn{2}{r}{2650 kg/m$^3$ $^*$}\\
filling factor $f$&\multicolumn{2}{r}{$0.36\pm 0.01$}\\
&&\\
\textbf{number of aggregates}&&\textbf{excited}\\
\quad total&$110$&$111$\\
&&\\
\multicolumn{2}{l}{\textbf{high speed}}&\textbf{excited}\\
rec. time&8175 ms&8175 ms\\
mean $\bar{v}_{col}$&4.0 mm/s&7.4 mm/s\\
collisions&348&1515\\
&\\
\multicolumn{2}{l}{\textbf{long term}}&\textbf{excited}\\
rec. time&900 sec.&900 sec.\\
collisions&$38.312^{**}$&$166.789^{**}$\\
\hline
\multicolumn{3}{l}{\footnotesize{$^{*}$constituent grains}}\\
\multicolumn{3}{l}{\footnotesize{$^{**}$extrapolated from high speed recording}}\\
\end{tabular}
\end{table}
The filling factor of $f=0.36$ is appropriate for compact dust aggregates
\citep{weidling2009,zsom2010,teiser2011b,meisner2012}. We note again, that our experiments do not model the collisions of the first mm-size aggregates which have
much lower filling factors. The experiments reported here model a somewhat later phase when mm-dust aggregates are already compacted by collisions. The average collision velocity $\bar{v}_{col}$ is shown in Tab.\ref{tab:coll} and the distribution of the collision velocities with and without excitation are shown in Fig.\ref{fig:vcoll1} and \ref{fig:vcoll2}.
A negative velocity means that the two aggregates move away from each other slowly while rotating and colliding with uneven parts of their rim.

Collisions were clearly visible in the high speed recording phase. In addition, the aggregates influence 
each other's trajectories by non-contact interactions. This might be interpreted as an effect of the gas outflow below the aggregates which influence we estimate below. 
As there is no physical contact between the aggregates during these interactions they are not counted as collision events.

To determine the collision details we tracked the position of all aggregates in every frame taken during the high speed recording phase. To automatically track the aggregates, the indiviual grayscale images are binarized: Background and aggregate pixel are seperated by their gray value. The experimental platform is black while the aggregates reflect some of the light and appear brighter in the grayscale images. To ensure best reults  we choose Otsu's algorithm \citep{otsu1979} for the binarization. The choice of the binarization threshold (i.e. the gray value to seperate background and aggregate) has an effect on the total collision number but almost no effect on the probabilities of the collisional outcomes (e.g. bouncing and sticking). Otsu's algorithm gives a relative conservative number of collisions compared to other binarization algorithms and hence the total number of collisions in the experiments is rather underestimated as overestimated.

The time step between the individual images is 5 ms. An interaction between two aggregates is defined as collision if the aggregates touch in at least one frame of the binarized image sequence. The boundary between sticking and bouncing is determined by two different timescales. One is the pure interaction time of colliding aggregates. This is the time two interacting aggregates need for a rebound which we estimate as follows: The minimum contact time for a bouncing collision can be estimated by elastic compression and relaxation of an aggregate by the other aggregate. With the modulus of elasticity $E$ the compression is defined as
\begin{equation}
\frac{F}{A}=-E\frac{(\Delta d)}{d}\label{eq:ela}
\end{equation}
with $F$ as repelling force, $A$ as contact area, $d$ and $\Delta d$ as diameter and change in diameter of the aggregate. The
force on the impinging aggregate is
\begin{equation}
F=m\cdot a=m\cdot \frac{d^2(\Delta d)}{dt^2}.\label{eq:force}
\end{equation}
where $m$ is the reduced mass. Combining Eq.(\ref{eq:ela}) and Eq.(\ref{eq:force}) results in
\begin{equation}
\frac{d^2(\Delta d)}{dt^2} =-\frac{EA}{md}(\Delta d)\label{eq:deltad}
\end{equation}
which is an harmonic oscillator and assuming $(\Delta d) = B \sin{\omega t}$ gives
\begin{equation}
\frac{d^2(\Delta d)}{dt^2} = -B \omega^2\sin{\omega t}.
\end{equation}
With Eq.(\ref{eq:deltad}) and the duration of a bouncing event $t_{bounce}$, which is half a period or $\omega \cdot t_{bounce} = \pi$, it is
\begin{equation}
t_{bounce} = \pi \left(\frac{EA}{md}\right)^{-1/2}.
\end{equation}
With $E\simeq 5\cdot 10^6$ Pa \citep{meisner2012}, $A=10^{-8}$ m$^2$, $m\simeq 10^{-7}$ kg and $d\simeq 1$ mm the estiamted bouncing time is $t_{bounce}\simeq 0.1$ ms. 

\begin{figure}
\centering
\includegraphics[width=84mm]{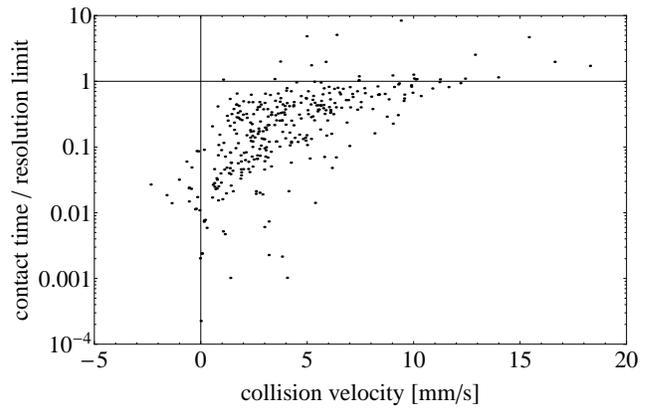}
\caption{Ratio of the contact time of aggregates to the resolution time over the collision velocity for the not excited case. For values below 1 no statement can be given if the event was sticking or bouncing. Values larger than 1 indicate sticking. The inset shows the complete data inculding the few extreme events.}
\label{fig:inter1}
\end{figure}

\begin{figure}
\centering
\includegraphics[width=84mm]{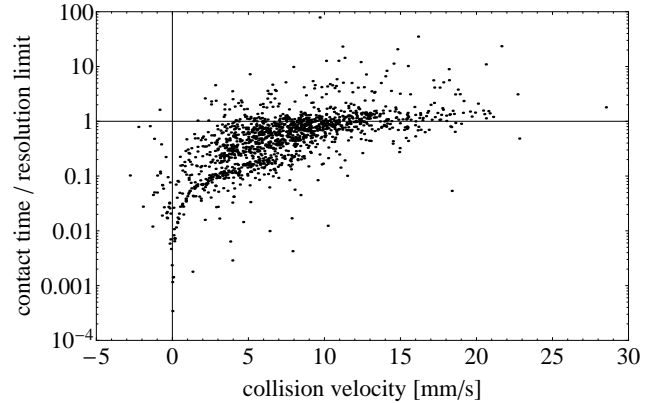}
\caption{Ratio of the contact time of joined aggregates to the resolution time over the collision velocity for the excited case. For values below one no statement can be given if the event was sticking or bouncing. The inset shows the complete data inculding the few extreme events.}
\label{fig:inter2}
\end{figure}

Important to decide if a collision lead to sticking is the timescale set by the spatial resolution of the observation which is $s = 46$ $\mu$m: The resolution time is therefore determined by the time the aggregates need to translate a length $4s$ ($2s$ for approach and $2s$ for moving apart again in the worst case). For the reproach the difference in the relative velocities before (index $b$) and after (index $a$) the interaction $C_I=v_{a}/v_{b}$ has to be considered. Within a time period of $t_{resolve} = 2s/v_{col}+2s/(C_I\cdot v_{col})$ two particles are observed to be in contact even for a perfect bouncing event. In Fig.\ref{fig:inter1} and Fig.\ref{fig:inter2} we show for all interaction events the ratio of the contact time (aggregates beeing visually in contact in the binarized images) to the individually calculated resolution time.

A value above one means that two aggregates are observed to be in contact longer then a potential bouncing event would last. For values below 1 no statement can be given if the interaction was a bouncing or a sticking event. In general, due to the resolution limit and the short timescales between successive collisions the experiments do not allow to classify a collision as bouncing. Values larger than 1 are a strong indication that the interaction was a sticking event. It has to be noted though that the reproach velocities are not rebound velocities. Therefore, we cannot rule out that e.g. two particles bounce off each other with low coefficient of restitution, are observed to be one aggregate for a longer time and reproach fast after a second collision occured. The resolution time calculated above is therefore a lower limit.  Strictly speaking the experiment does not allow to quantify this further. To judge this, well separated collisions observable undisturbed for a sufficiently long time are needed. That a collision is bouncing gets less likely the larger the ratio between observed lifetime and resolution time.

For the not excited case $93.3${\%} of the collisions the ratio of the contact times to resolution times are below one, $6.7${\%} are above one and $1.4${\%} are above two. The maximum ratio is $8$. For the excited case $76.4${\%} are below one, $23.6${\%} are above one and $5.8${\%} are above two. The maximum ratio here is $79$. 

Beyond giving exact numbers for individual collisions, which is not the focus of this paper, there are sticking collisions (i.e. the ratios of contact times to resolution limit above 2). Proof of this are also aggregates which are not only in contact but also show bound rotation. 

\begin{figure}
\centering
\includegraphics[width=84mm]{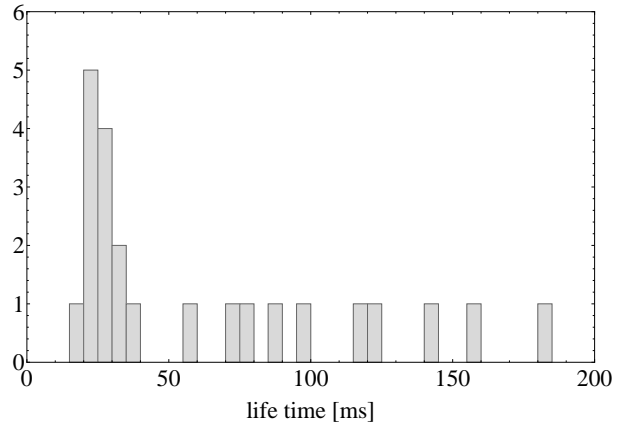}
\caption{Distribution of the lifetimes of grown aggregates (highspeed recording phase, no excitation).}
\label{fig:lifetimes}
\end{figure}

\begin{figure}
\centering
\includegraphics[width=84mm]{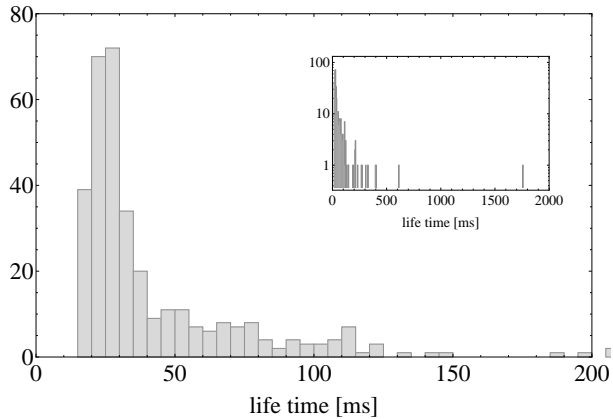}
\caption{Lifetimes of aggregates, same as Fig. \ref{fig:lifetimes} but for the excited system. The inset shows the complete data including the few extreme lifetimes.}
\label{fig:lifetimes2}
\end{figure}

We classifiy the collisions with a ratio above one as aggregates stuck together. The observed distribution of the lifetimes of these grown aggregates are shown in Fig.\ref{fig:lifetimes} and Fig.\ref{fig:lifetimes2}.

\begin{figure}
\centering
\includegraphics[width=84mm]{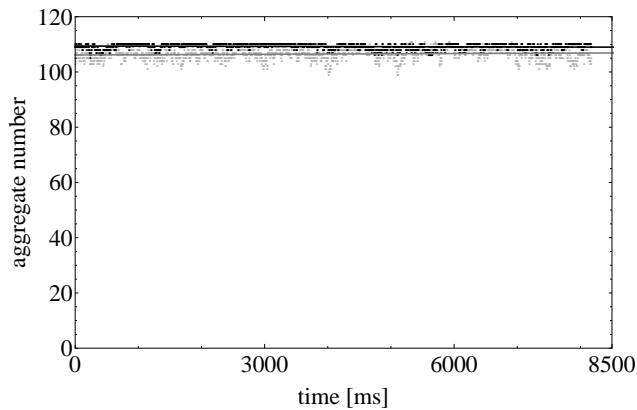}
\caption{Total number of aggregates during the high speed recording phase (dark grey: no excitation; light grey: with excitation). The solid lines are linear fits to the data.}
\label{fig:num1}
\end{figure}

\begin{figure}
\centering
\includegraphics[width=84mm]{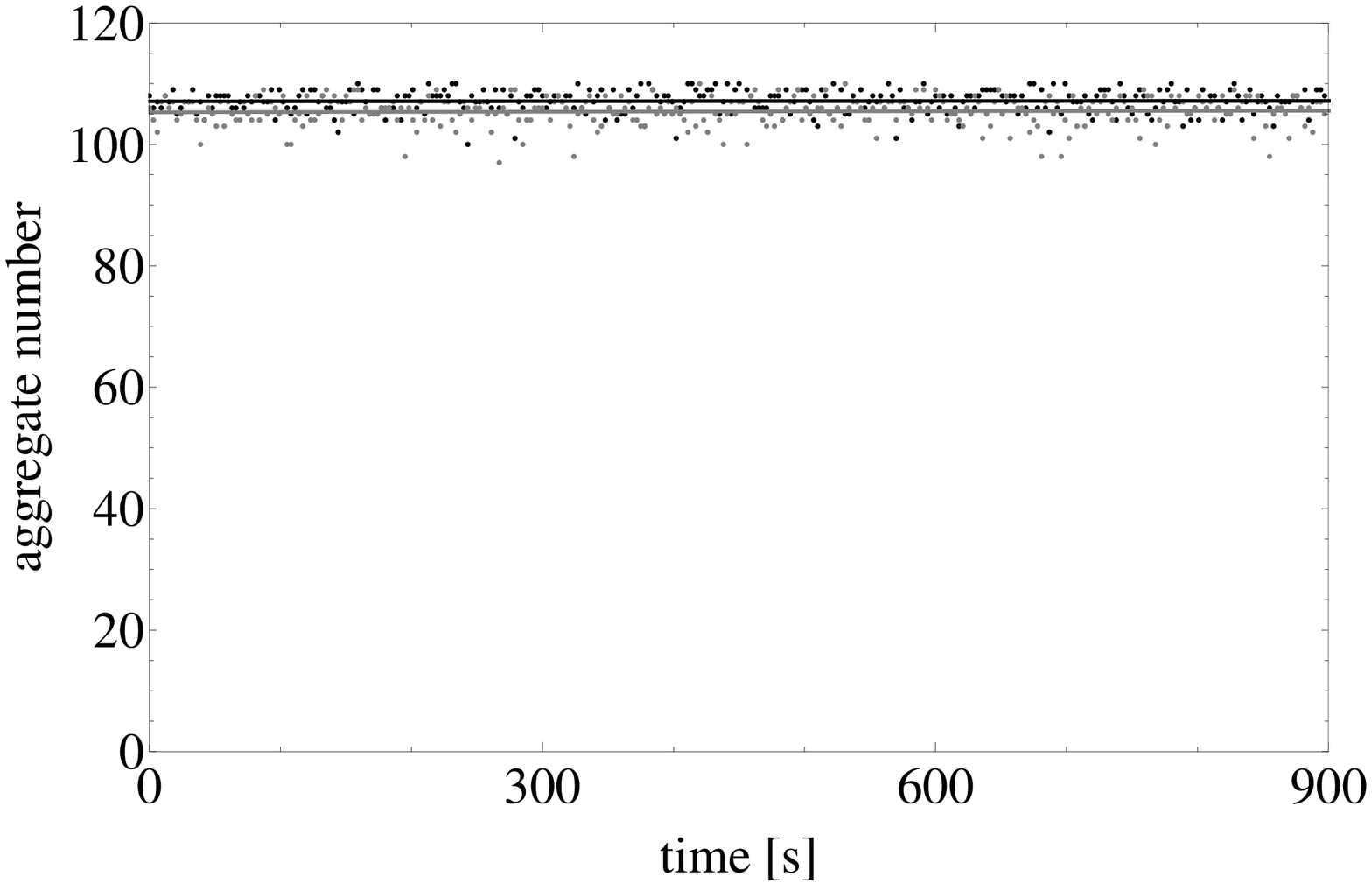}
\caption{Total number of aggregates during the long time recording phase (dark grey: no excitation; light grey: with excitation). The solid lines are linear fits to the data.}
\label{fig:num2}
\end{figure}

The analysis so far concentrated on individual collisions. To estimate the temporal evolution of the aggregate ensemble we observed the
particle ensemble for 15 minutes, taking one image every 3 seconds. In case that growth would be efficient the number of aggregates should decrease from 110 to smaller values. Fig.\ref{fig:num1} shows the total number of aggregates  during the high speed recording phase and Fig.\ref{fig:num2} shows the total number of aggregates over the whole recording period of 15 minutes.

As our experimental platform is slightly concave and as there are non contact forces between approaching aggregates (Knudsen compressor gas outflow below the aggregates) we estimate the strength of these effects. The force occuring during a pure rebound is given by Eq.(\ref{eq:ela}) and results in $F_{reb._1} = 9\times 10^{-6}$ N for the not excited case and $F_{reb._2}=2\times 10^{-5}$ N for the excited case. The slightly concave platform has a radius of 30 mm and is 0.8 mm deep at maximum (center of platform). The residual gravitational force acting on the aggregates pulls them together. In the worst case this force is $F_{res.G}=4\times 10^{-9}$ N. The deceleration of two approaching aggregates determines the non contact pushing force (gas outflow because of the Knudsen compressor effect) which takes the value of $F_{Kn} \simeq 10^{-11}$ N. We observe in our experiments that most grown aggregates break apart again without having contact to other aggregates ($95.7${\%} in the not excited case and $81.5${\%} in the excited case). Hence, the non contact forces (streaming gas, laboratory gravity), which are orders of magnitudes weaker than the forces occuring during a real collision event, are strong enough to break grown aggregates. In other words: If these weak non contact forces are strong enough to break aggregates apart, any real collision would break grown aggregates for certain.

The majority of the individual collisions show a ratio of the contact time to the resolution limit well below one and hence no quantitative number number can be given what fraction of the interaction was a sticking or bouncing event. However, the exact number is of minor importance here. We do
see a significant number of clear sticking events in individual interactions. Nevertheless, all grown aggregates break apart again. During the time scale of the experimental procedere no stable net growth is observed. Expressed differently, after a time period of 15 minutes with more than $10^5$ collisions or more than $1000$ collisions per aggregate no net growth is visible.

\section{Application to protoplanetary discs}
\label{sec:app}
All the aggregates in the experiment are moving freely. We do not have any evidence of contacts with the surface of the heater. There are additional forces (streaming gas, laboratory gravity) acting on the aggregates in the laboratory which are not present in a protoplanerary disc. However, these forces are orders of magnitudes weaker than the force applied on an aggregate during an actual collision. If these forces are sufficient to break aggregates again any real collision would break an aggregate as well (see discussion of the end of section 3). Therefore, we argue that the experiment is well suited to simulate collisions in protoplanetary discs. Also, as the collision velocity distributions do not depend on time they very well simulate the analog situation of the disc, where collision velocities stay the same as the particles have enough time to couple to the gas in between two collisions. The most striking observation is that we do not observe a stable growth of a larger particle although sticking in individual collisions appears. Even after 1000 collisions the individual aggregates are still individual aggregates. This is a perfect illustration of a bouncing barrier as proposed by \citet{zsom2010} and as found to be beneficial for further evolution by \citet{windmark2012}.

On average each particle collides $1500$ times with other particles but does not grow (excited case). Compared to protoplanetary discs this would simulate a time span of $1\times 10^6$ years at 1 AU in a minimum mass solar nebula estimated as follows.

The gas density at 1 AU is about  $10^{-6}$ kg/m$^3$ \citep{hayashi1985}. Assuming a dust to gas ratio of $1/100$ the dust density is about $10^{-8}$ kg/m$^3$. If all mass would be in mm-particles (volume $10^{-9}$ m$^3$) with a density of 1000 kg/m$^3$ (low density aggregates) one particle has a mass of $10^{-6}$ kg. This equals a particle number density of $n = 10^{-8} / 10^{-6}$ m$^{-3}$ or $10^{-2}$ /m$^3$.  The time scale for a collision is $\tau = 1 / (n v \sigma)$ with $\sigma$ being the particle cross section ($4\times 10^{-6}$ m$^2$) and $v = 10^{-3}$ m/s. This yields $\tau \simeq 3\times10^{10}$ s  or approx. $1000$ years.  Hence, $1500$ collisions relate to about $1\times 10^6$ years in the disc (for the specific location, density and velocity). Therefore, the bouncing barrier is very persistent even if further collisions in the experiments would eventually lead to the formation of larger aggregates. The latter is not expected though. As there is no sign of growth in the experiments the bouncing barrier can be considered as very robust for the total duration of the disc lifetime.

The experiment only considered mm-aggregates of one composition and grain size, one filling factor of 36 \%, and a specific velocity distribution.  This might be different in protoplanetary discs. At what particle size the barrier is actually located in a disc,
depending on the radial distance to the star, is subject to future work.

\section{Conclusions}
\label{sec:conlusions}

Millimeter size dust aggregates are observed in protoplanetary discs \citep{rodmann2005}. The accepted picture is that they grow
by sticking collisions \citep{blum2008}. The naive picture of a continuous particle growth by hit-and-stick beyond the
size of mm or cm is certainly missing key points. Often discussed is the fragmentation barrier at meter size. This means that particles 
which encounter large meter size bodies do so with high speed of 50 to 60 m/s and they might loose their integrity and not just stick
to the meter body. More recently \citet{zsom2010} explicitly introduced the term bouncing barrier for compacted mm size particles. Here the idea is that collisions of mm dust aggregates among each other only lead to bouncing. While this is not undoing earlier growth it does
prevent further growth if no sticking occurs. 

However, both processes -- bouncing at small size and fragmentation at large size -- might
be beneficial for growth. Laboratory experiments by \citet{wurm2005}, \citet{teiser2009b} and Meisner et al. (accepted) show that 
fragmentation at collision velocities up to 60 m/s do not necessarily destroy a large target body. While large parts of the projectile can
be destroyed, this dissipation of kinetic energy still allows a fraction of several ten \% of the projectile dust to stick to a larger
target. \citet{windmark2012} showed that growth of larger seeds is possible if a bouncing barrier prevents all particles to grow
at the same time. Collisional growth beyond the barriers is therefore possible if seeds are provided but the bouncing barrier in general is
very important in that context.

In this paper we studied the long term collisional behaviour of a set of about 100 equal dust aggregates of about 1 mm in size at mm/s
collision velocities. Individual collisions show  some sticking effeciency but a dimer of two aggregates cannot just be
regarded as a new particle with different size as the newly formed connection is rather weak compared to the internal structure of
each of the two individual aggregates. \citet{jankowski2012} already showed that detachment in a further collision produces
the original aggregates, eventually, even though \citet{jankowski2012} and \citet{weidling2012} also find further attachments. So far these were only individual collisions though. Here we study a total of over $10^5$ collisions
for 100 particles or 1000 collisions per mm-aggregate. We find sticking and detachment but no large aggregate that survived
for long. In the end no net growth was found. This is a perfect demonstration of a bouncing barrier though in our experiments it might be better called a detachment barrier.

The parameters of the dust -- $\mu m$ grain size, mm aggregate size, mm/s collision velocities -- are close to values expected 
in protoplanetary discs. Applied to a disc our simulation experiment would suggest that -- without a seed -- no further growth would
be possible over the discs lifetime in the terrestrial planet forming region. If seeded growth or gravitational instabilities
\citep{chiang2010} feeding on the mm or cm particles would kick in to form planetesimals remains to be seen.
However, a bouncing (or detachment) barrier at mm to cm size is very likely an important milestone in planetesimal formation.

\section*{Acknowledgements}
We acknowledge funding by the DFG as part of the research group FOR 759 and
the project Ke 1897/1-1. We thank the anonymous reviewer for the comments.

\label{lastpage}


\begin{thebibliography}{}
\bibitem[\protect\citeauthoryear{Aumatell \& Wurm}%
{2011}]{aumatell2011} Aumatell G., Wurm G., 2011, MNRAS, 418, L1

\bibitem[\protect\citeauthoryear{Ayliffe et al.}%
{2012}]{ayliffe2012} Ayliffe B. A., Laibe G., Price D. J., Bate M. R., 2012, MNRAS, 423, 1450

\bibitem[\protect\citeauthoryear{Barge \& Sommeria}%
{1995}]{barge1995} Barge P., Sommeria J., 1995, A{\&}A, 295, L1

\bibitem[\protect\citeauthoryear{Beitz et al.}%
{2011}]{beitz2011} Beitz E., G{\"u}ttler C., Blum J., Meisner T., Teiser J., Wurm G. 2011, ApJ, 736, 34

\bibitem[\protect\citeauthoryear{Birnstiel et al.}%
{2012}]{birnstiel2012} Birnstiel T., Klahr H., Ercolano B., 2012, A{\&}A, 539, A148

\bibitem[\protect\citeauthoryear{Blum \& Wurm}%
{2000}]{blum2000} Blum G., Wurm G., 2000, Icarus, 143, 138

\bibitem[\protect\citeauthoryear{Blum \& Wurm}%
{2008}]{blum2008} Blum G., Wurm G., 2008, ARA{\&}A, 46, 21

\bibitem[\protect\citeauthoryear{Blum et al.}%
{2000}]{blum2000b} Blum J. et al., 2000, Phys. Rev. Lett., 85, 2426

\bibitem[\protect\citeauthoryear{Bodrova et al.}%
{2012}]{bodrova2012} Bodrova A., Schmidt J., Spahn F., Brilliantov N., 2012, Icarus, 218, 60

\bibitem[\protect\citeauthoryear{Brauer et al.}%
{2008}]{brauer2008} Brauer F., Henning T., Dullemond C. P., 2008, A{\&}A, 487, L1

\bibitem[\protect\citeauthoryear{Chiang \& Youdin}%
{2010}]{chiang2010} Chiang E., Youdin A. N., Wurm G., 2010, Annu. Rev. Earth Planet. Sci., 38, 493

\bibitem[\protect\citeauthoryear{Chokshi et al.}%
{1993}]{chokshi1993} Chokshi A., Tielens A. G. G. M., Hollenbach D., 1993, ApJ, 407, 806


\bibitem[\protect\citeauthoryear{Colwell}%
{2003}]{colwell2003} Colwell J. E. et al., 2003, Icarus, 164, 188

\bibitem[\protect\citeauthoryear{Colwell et al.}%
{2008}]{colwell2008} Colwell J. et al, 2008, Icarus, 195, 908

\bibitem[\protect\citeauthoryear{Cuzzi et al.}%
{2008}]{cuzzi2008} Cuzzi J. N., Hogan R. C., Shariff K., 2008, ApJ., 687, 1432

\bibitem[\protect\citeauthoryear{Deckers \& Teiser}%
{2013}]{deckers2013} Deckers J., Teiser J., 2013, ApJ, 769, 151

\bibitem[\protect\citeauthoryear{Dittrich et al.}%
{2013}]{dittrich2013} Dittrich K., Klahr H., Johansen A., 2013, ApJ, 763, 117

\bibitem[\protect\citeauthoryear{Dominik \& Tielens}%
{1997}]{dominik1997} Dominik C., Tielens A. G. G. M., 1997, ApJ, 480, 647

\bibitem[\protect\citeauthoryear{Hayashi et al.}%
{1985}]{hayashi1985} Hayashi C., Nakazawa K., Nakagawa Y., 1985, Protostars and Planets II, 1100

\bibitem[\protect\citeauthoryear{Hei{\ss}elmann et al.}%
{2010}]{heisselmann2010} Hei{\ss}elmann D., Blum J., Fraser H. J., Wolling K., 2010, Icarus, 206, 424

\bibitem[\protect\citeauthoryear{Higa et al.}%
{1998}]{higa1998} Higa M., Arakawa M., Maeno N., 1998, Icarus, 133, 310

\bibitem[\protect\citeauthoryear{Jankowski et al.}%
{2012}]{jankowski2012} Jankowski T., Wurm G., Kelling T., Teiser J., Sabolo W., Guti\'{e}rrez P. J., Bertini I., 2012, A{\&}A, 542, 80


\bibitem[\protect\citeauthoryear{Johansen et al.}%
{2007}]{johansen2007} Johansen A., Oishi J. S., Mac Low M. M., Klahr H., Henning T., Youdin A., 2007, Nature, 448, 1022


\bibitem[\protect\citeauthoryear{Kelling \& Wurm}%
{2009}]{kelling2009} Kelling T., Wurm G., 2009, Phys. Rev. Lett., 103, 215502--1


\bibitem[\protect\citeauthoryear{Kelling et al.}%
{2011}]{kelling2011c} Kelling T., Wurm G., D\"urmann C., 2011, Rev. Sci. Instrum., 82, 115105

\bibitem[\protect\citeauthoryear{Klahr \& Henning}%
{1997}]{klahr1997} Klahr H. H., Henning T., 1997, Icarus, 128, 213

\bibitem[\protect\citeauthoryear{Knudsen}%
{1909}]{knudsen1909} Knudsen M., 1909, Ann. Phys., 336, 633

\bibitem[\protect\citeauthoryear{Konopka et al.}%
{2005}]{konopka2005} Konopka U. et al., 2005, New J. Phys., 7, 227

\bibitem[\protect\citeauthoryear{Kothe et al.}%
{2013}]{kothe2013} Kothe S., Blum J., Weidling R., G{\"u}ttler C., 2013, Icarus, 225, 75

\bibitem[\protect\citeauthoryear{Kothe et al.}%
{2010}]{kothe2010} Kothe S., G{\"u}ttler C., Blum J., 2010, ApJ, 725, 1242


\bibitem[\protect\citeauthoryear{Meisner et al.}%
{2012}]{meisner2012} Meisner T., Wurm G., Teiser J., 2012, A{\&}A, 544, A138


\bibitem[\protect\citeauthoryear{Muntz et al.}%
{2002}]{muntz2002} Muntz E. P., Sone Y., Aoki K., Vargo S., Young M., 2002, J. Vac. Sci. Technol., A, 20, 214

\bibitem[\protect\citeauthoryear{Okuzumi}%
{2009}]{okuzumi2009} Okuzumi S.,2009, ApJ., 698, 1122

\bibitem[\protect\citeauthoryear{Okuzumi et al.}%
{2012}]{okuzumi2012} Okuzumi S., Tanaka H., Kobayashi H., Wada K., 2012, ApJ., 752, 106

\bibitem[\protect\citeauthoryear{Otsu}%
{1979}]{otsu1979} Otsu N.,1979, IEEE Trans. System., Man and Cybernetics, 9, 62

\bibitem[\protect\citeauthoryear{Paszun \& Dominik}%
{2006}]{paszun2006} Paszun D., Dominik C., 2006, Icarus, 182, 274

\bibitem[\protect\citeauthoryear{Pinilla et al.}%
{2012}]{pinilla2012} Pinilla P., Birnstiel T., Ricci L., Dullemond C. P., Uribe A. L., Testi L., Natta A., 2012, A{\&}A, 538, A114

\bibitem[\protect\citeauthoryear{Rodmann et al.}%
{2005}]{rodmann2005} Rodmann J., Henning T., Chandler C. J., Mundy L. G., Wilner D. J., 2005, A{\&}A., 446, 211

\bibitem[\protect\citeauthoryear{Ros \& Johansen}%
{2013}]{ros2013} Ros K., Johansen A., 2013, A{\&}A, 552, A137

\bibitem[\protect\citeauthoryear{Saito \& Sirono}%
{2011}]{saito2011} Saito E., Sirono S. I., 2011, ApJ, 728, 20

\bibitem[\protect\citeauthoryear{Schr{\"a}fer et al.}%
{2007}]{schaefer2007} Schr{\"a}fer C., Speith R., Kley W.,  2007, A{\&}A, 470, 733

\bibitem[\protect\citeauthoryear{Schr{\"a}pler et al.}%
{2012}]{schraepler2012} Schr{\"a}pler R., Blum J., Seizinger A., Kley W., Wilner D. J., 2012, ApJ., 758, 35

\bibitem[\protect\citeauthoryear{Seizinger \& Kley}%
{2013}]{seizinger2013} Seizinger A., Kley W., 2013, A{\&}A, 551, A65

\bibitem[\protect\citeauthoryear{Sirono}%
{2011}]{sirono2011} Sirono A., 2011, ApJ, 735, 131

\bibitem[\protect\citeauthoryear{Supulver et al.}%
{1995}]{supulver1995} Supulver K. D., Bridges F. G., Lin D. N. C., 1995, Icarus, 113, 188


\bibitem[\protect\citeauthoryear{Teiser et al.}%
{2011a}]{teiser2011} Teiser J., K{\"u}pper M., Wurm G., 2011, Icarus, 215, 596

\bibitem[\protect\citeauthoryear{Teiser et al.}%
{2011b}]{teiser2011b} Teiser J., Engelhardt I., Wurm G., 2011, Apj, 741, 5

\bibitem[\protect\citeauthoryear{Teiser \& Wurm}%
{2009}]{teiser2009b} Teiser J., Wurm G., 2009, MNRAS, 393, 1584

\bibitem[\protect\citeauthoryear{van Eymeren \& Wurm}%
{2012}]{vaneymeren2012} van Eymeren J., Wurm G., 2012, MNRAS, 420, 183

\bibitem[\protect\citeauthoryear{Wada et al.}%
{2009}]{wada2009} Wada K., Tanaka H., Suyama T., Kimura H., Yamamoto T., 2009, ApJ., 702, 1490

\bibitem[\protect\citeauthoryear{Wada et al.}%
{2011}]{wada2011} Wada K., Tanaka H., Suyama T., Kimura H., Yamamoto T., 2011, ApJ., 737, 36

\bibitem[\protect\citeauthoryear{Weidenschilling}%
{1977}]{weidenschilling1977} Weidenschilling S. J., 1977, MNRAS, 180, 57

\bibitem[\protect\citeauthoryear{Weidling et al.}%
{2009}]{weidling2009} Weidling R., G{\"u}ttler C., Blum J., Brauer, F., 2009, ApJ., 696, 2036

\bibitem[\protect\citeauthoryear{Weidling et al.}%
{2012}]{weidling2012} Weidling R., G{\"u}ttler C., Blum J., 2012, Icarus, 218, 688

\bibitem[\protect\citeauthoryear{Windmark et al.}%
{2012a}]{windmark2012} Windmark F., Birnstiel T., G{\"u}ttler C., Blum J., Dullemond C. P., Henning T., 2012a, A{\&}A, 540, A73

\bibitem[\protect\citeauthoryear{Windmark et al.}%
{2012b}]{windmark2012b} Windmark F., Birnstiel T., Ormel C. W., Dullemond C. P., 2012b, A{\&}A, 544, L16

\bibitem[\protect\citeauthoryear{Wurm \& Blum}%
{1998}]{wurm1998} Wurm G., Blum J., 1998, Icarus, 132, 125

\bibitem[\protect\citeauthoryear{Wurm et al.}%
{2005}]{wurm2005} Wurm G., Paraskov G., Krauss O., 2005, Icarus, 178, 253

\bibitem[\protect\citeauthoryear{Zsom et al.}%
{2010}]{zsom2010} Zsom A., Ormel C. W., G{\"u}ttler C., Blum J., Dullemond C. P., 2010, A{\&}A, 513, 56

\end{thebibliography}
\end{document}